# Tunable transmissive metasurface based on thin-film lithium niobate

*Zetian Chen\*, Noa Mazurski, Jacob Engelberg, and Uriel Levy\**

*Corresponding authors: Zetian Chen, Uriel Levy*

Institute of Applied Physics, The Faculty of Science, The Center for Nanoscience and Nanotechnology, The Hebrew University of Jerusalem

## ABSTRACT

Active metasurfaces hold great promise for spatial light modulation, and electro-optic tuning using lithium niobate is particularly attractive due to its high transparency and well-established thin-film platform. In this work, we present a free-space transmissive light modulator based on a seemingly un-patterned thin-film lithium niobate-on-insulator platform, integrated with a transparent conductive oxide meta-grating fabricated through a single lithography process. Guided mode resonance in the near-infrared region is induced with high mode confinement within lithium niobate layer, which directly in contact with the electrodes. A notable resonance shift of 0.38 nm is observed for the fundamental mode under ±10 V bias, while a maximum modulation amplitude of 4.6% is achieved for another higher-order mode. Incident angle is further exploited as another tuning parameter to split and sensitively shift the resonances. These results demonstrate the potential of this design for applications in compact, scalable, and tunable spatial light modulation devices.

# Introduction

Metasurfaces, composed of engineered planar periodic subwavelength structures, have been proven to be promising candidates for spatial light modulations[1–4]. While traditional metasurfaces are typically passive, recent advancements have introduced active metasurfaces[5], which can dynamically alter their optical properties in response to external stimuli like electrical[6], thermal[7], or optical signals[8]. These active metasurfaces enable real-time control of light, facilitating applications in dynamic beam steering[9,10], tunable lenses[11,12], reconfigurable holography[13], display[14], and adaptive optics for communication and sensing technologies[15]. Among the various control approaches, electrically tunable metasurfaces through the electro-optic (EO) Pockels effect are particularly attractive due to their linearity, potentially fast response, and ability to integrate with electronic systems[16–21]. Various materials have been explored for EO modulation, primarily solution-processed EO organic polymers [18,22–24] where pre-poling is usually required before use, and non-centrosymmetric crystals, having strong Pockles nonlinearity. Lithium niobate (LNO) stands out among the latter category for active photonic applications due to its high EO coefficient, wide transparent window, and excellent chemical and mechanical stability[25–27]. Moreover, with the maturation of thin-film lithium niobate on insulator (LNOI) technology[28], devices, both integrated[29,30] and free-space[31–39], can now be fabricated using standard and reproducible micro/nanofabrication processes, making them highly attractive for future commercialization. However, direct etching of LNO with high-quality profile remains a challenge[40,41]. Besides, products like thin-film LNO on transparent conductive oxide (TCO) are so far not available, hindering the realization of novel designs[31,42,43] relying on such platforms.

Here, we have designed, fabricated and experimentally demonstrated a free-space light chip scale amplitude modulation device based on an effectively etch-less thin-film LNO and Indium tin oxide (ITO) meta-grating that operates in transmission mode. The device functions at a design wavelength of 970 nm within the near-infrared (NIR) region (750 nm–1000 nm), featuring high background transmittance and high-quality factor (Q-factor of 440) guided mode resonances (GMRs) at the design wavelength. We systematically investigate the variation in EO tuning efficiency for different resonance modes using both experiments and simulations. Upon the application of $\pm 10$ V bias, a clear spectrum shift of 0.38 nm is observed for the designed wavelength, with a maximum modulation amplitude of 4.6% realized for a higher order mode within NIR. Furthermore, we demonstrate that the incident angle serves as an additional post-fabrication tuning parameter, in combination with EO modulation. The novel design presented in this work holds promise for future efficient and compact transmission light EO modulation applications.

**Results**

*The sample*

The device structure and its working principle are depicted in Fig. 1. As shown in the cross-sectional view of one unit cell in Fig. 1(a), the device is built on a commercially available x-cut lithium niobate on insulator (LNOI) platform with 310 nm thick LNO on ~2 μm of $SiO_2$. ITO nanobars (t=100 nm) are deposited and patterned on top of the LNO, with shallow over-etching into LNO due to the fabrication process as will be clarified further below. Each unit cell consists of a pair of bars positioned in close proximity at a distance of d = 220 nm. The device is additionally encapsulated with a layer of PMMA to prevent air breakdown upon applying bias. In theory, this hybrid waveguide structure—comprising ITO nanobars atop the thin-film LNO—can support a fundamental TE mode within the xy-plane at ~970 nm, as depicted in Fig. 1(b). The figure shows the electric field intensity profile obtained from eigen mode analysis for two adjacent unit cells. A vector plot of the mode's x-polarization is provided in Fig. S1. Owing to the higher refractive index of LNO compared to ITO, along with the special arrangement of the bars, the guided mode is largely confined within the LNO region, with 81% of the mode power concentrated there, especially beneath the adjacent nanobars. Besides electrical conductivity, ITO is chosen for its low optical loss in the NIR region (with its refractive index n and extinction coefficient k shown in Fig. S2). When a bias is applied through the two ITO nanobars in each unit cell, an electric field, schematically shown as black arrow lines in Fig. 1(b), develops inside the LNO where the mode resides. This field induces a refractive index change in the LNO via the electro-optic (EO) Pockles effect. To maximize the EO effect by utilizing the largest EO coefficient ($r_{33}$), we align the extraordinary optical axis of the LNO with the mode polarization direction[44]. Consequently, in the area around the center of the bars (indicated by the white dashed lines), where the mode is most concentrated and the electric field has a strong x-directional component, the change in refractive index can be expressed as

$$\Delta n_e = -\frac{1}{2} n_e^3 r_{33} E_x \qquad (1)$$

where $n_e$ is the refractive index along the extraordinary axis, $E_x$ is the electrical field component in the x-direction, and $r_{33} \approx 30$ pm/V is the EO coefficient for this configuration. Although there is an opposing field between the bars of neighboring cells, it is much weaker due to their larger separation, and also because the optical mode is largely absent in these areas. Therefore, the EO effect contributes collectively among the neighboring cells.

To couple light into such mode from normal incident light coming from free space, we create a periodic notch at the outer side of each bar to compensate for the momentum mismatch[45], as illustrated in the top view of the device in Fig. 1 (c). For a given notch period (p) with sufficiently long bars, multiple orders of guided mode resonances (GMR) can be excited across a broad spectral range according to equation (2):

$$p = \lambda_{eff,i} = \lambda_0/n_{eff,i} \tag{2}$$

where $\lambda_{eff,i}$ is the effective guided mode wavelength for i$^{th}$ order GMR, $n_{eff,i}$ is its effective refractive index, and $\lambda_0$ is the free-space wavelength. A notch period p=505 is chosen, as the calculated n$_{eff}$=1.925 for the fundamental mode (Fig. 1 (b)) at 970 nm. Because a relatively large notch is used (see Fig. 1 (c), p'/p≈0.4, w'/w≈0.8, with w=380 nm), for reasons explained later, we propose a transformation to convert the geometry with broken translational symmetry (along the y-axis) into a translationally symmetrical one, as illustrated in Fig. S3. To simplify the calculation, this translational symmetrical structure is used for eigen mode and electrical field simulation. The fabricated structure is shown in Fig.1 (d) and the inset of Fig. 1 (c) (top view), with unit cell length l=2200 nm. The ITO nanobars are etched by Ar etching[46] with metal mask defined by lift-off, as detailed in Method and Fig. S4. Energy-dispersive X-ray spectroscopy (EDS) element mapping of Indium shown in Fig. S5 (a) confirms that the ITO is etched cleanly between adjacent bars. This is crucial to ensure the proper establishment of the electric field between the bars, avoiding potential thermal effects caused by short circuits. A slight over-etch of about 50 nm into the LNO is observed, with a smaller over-etch between the closely spaced bars (see the zoomed-in view in Fig. S5(b)). This over-etch is unintentional and is primarily caused by the metal mask removal process. However, as demonstrated by the mode profile simulation in Fig. S6 for a device without any over-etch, this slight over-etch improves the confinement of the mode within each unit cell.

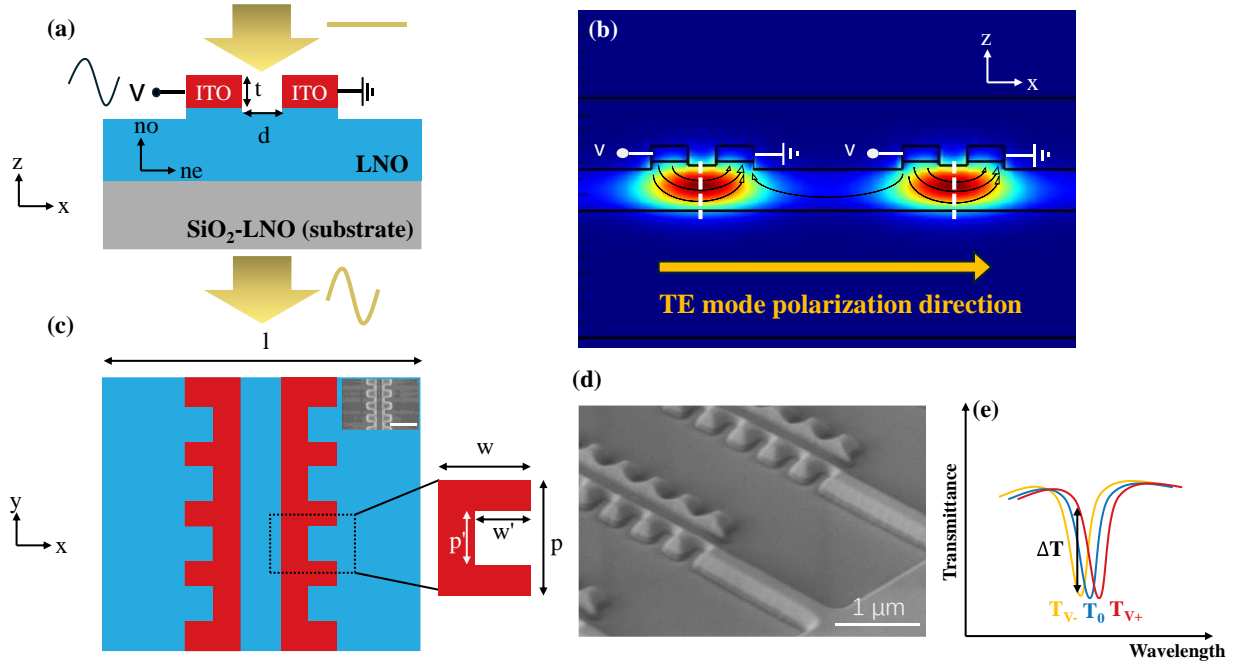

*Fig.1 Device structure and working principle. (a) Device cross section. (b) Fundamental TE guided mode electrical field intensity profile of two unit-cells that the structure supports at 970 nm. The mode is confined in each unit cell beneath the adjacent bars. White dashes lines: middle of the bar-pair where the mode is most concentrated, and electrical field aligns with the mode direction as well as the extraordinary optical axis of LNO, so that the largest EO coefficient can be utilized. (c) Device top view. It is based on a LNO on insulator platform, with a pair of adjacent periodically notched ITO nanobars in each unit cell forming meta-grating in x-direction. x-polarized light normally incidents on the metasurface and couples to in-plane guided mode through guided mode resonance. The electrical field applied through the bars changes the effective refractive index of the modes in the LNO by the EO effect and modulates the transmitted light. (d) and inset of (c) Tilted 55° and top view SEM images of a fabricated device. The white scale bars represent 1 μm for both. The images were taken before PMMA coating. (e) Schematics of the device functionality as a transmitted light modulator. Depending on the polarity of the electrical field, the change in refractive index will shift the spectrum in either direction around resonance. Probing the transmittance at a certain frequency can monitor the electrical modulation. Device dimensions: t=100 nm, d=220 nm, l=2200 nm, p=505 nm, w=380nm, w'/w=0.8, p'/p=0.4.*

Fig. 1(e) illustrates the functionality of the device studied. It functions as a light amplitude modulator operating in transmission mode. By employing the periodic notch structure, the device exhibits guided mode resonance (GMR) with a high-quality factor (Q) at the designed wavelength, which manifests as a transmission dip. When an electric field is applied across the ITO nanobars, the effective refractive index of the mode near the resonance is altered through the electro-optic (EO) effect. This results in either a blue or red shift of the resonance curve, depending on the polarity of the applied field. Thus, by monitoring the transmission power at a fixed wavelength around the resonance, the intensity of the transmitted light can be effectively modulated by the electric field.

*Transmission Spectrum*

Initially, the device static (no voltage) transmission spectrum was characterized and compared to the calculated results. A supercontinuum broadband laser, paired with a monochromator, was used as a tunable light source. The single-mode light beam, linearly polarized along the x-direction, was weakly focused onto the device using a long focal length lens to approximate incidence with a nearly collimated plane wave (as detailed in Fig.S7 and Method). By sweeping the monochromator cross a wide range of wavelengths (750 nm~990 nm), the transmission spectrum can be measured, allowing the capturing and comparison of multiple orders of GMR. Fig. 2(a) and (b) show the measured and calculated spectra of two devices with different notch periods (p = 505 nm and p = 500 nm). Two prominent resonances were observed within the measured wavelength range: the designed resonance around ~970 nm, and another high-contrast GMR occurring at ~775 nm, which may also be used for modulation. Additionally, a weaker resonance appears around ~820 nm (see Fig. S8), and the full spectrum of a single device is provided in Fig. S9. The experimental results are in good agreement with the simulated spectra in terms of the resonance wavelengths and the blue shift observed for the device with the shorter notch period. According to equation (2), this shift confirms that the resonance is due to the guided mode. The resonances exhibit a slightly asymmetrical shape, which is most likely due to the asymmetry in refractive indexes of guided modes claddings[47]. It should be noted that the spectrum is measured as absolute transmittance, using air as reference. The device demonstrates high transmission (~70 % in experiments, and >90 % in simulation at ~970 nm) in the resonance-free range, showcasing the advantage of ITO as a highly transparent meta-grating. We achieved high-Q GMRs, with a Q-factor of 440 at 968.5 nm and 270 at 777.5 nm. The corresponding simulated Q-factors were 922 and 300, respectively. The lower Q-factors and shallower contrast observed in the experimental results are likely due to fabrication imperfections and the finite illumination area.

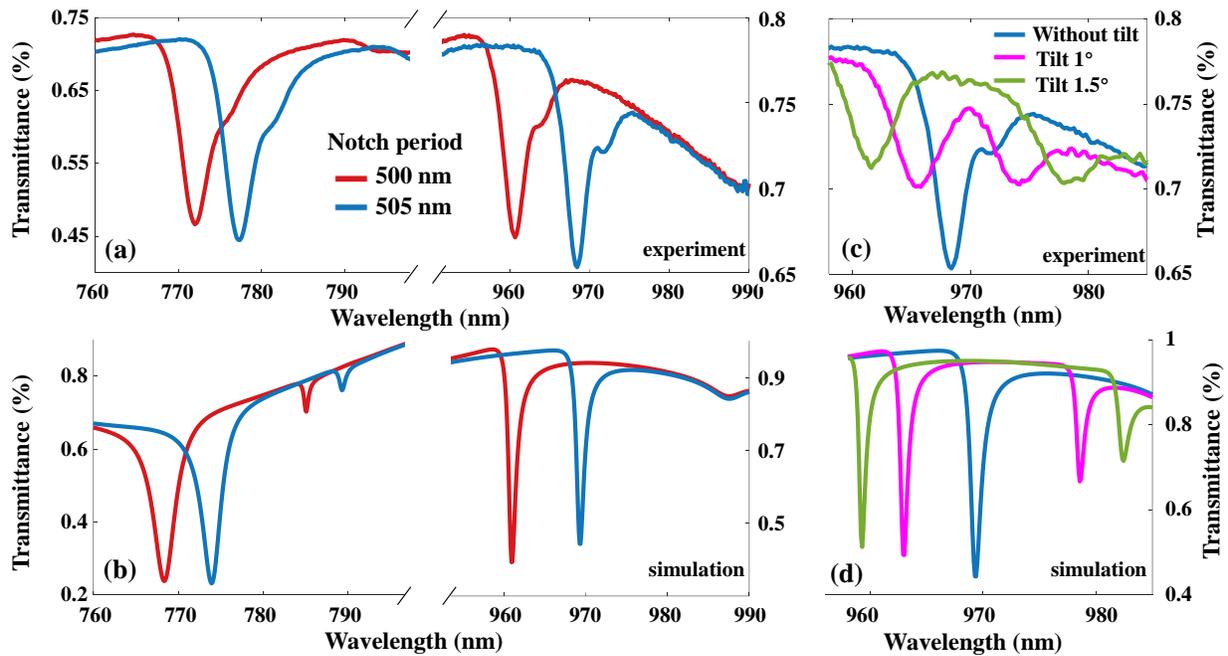

*Fig. 2 Device guided mode resonance (GMR) characterization. (a) and (b) Experimental and simulated transmittance spectrum of two GMRs at ~775 nm and ~970 nm, with different notch periods of p=500 nm and p=505 nm. (c) and (d) Experimental and simulated transmittance spectrum around the fundamental mode at*

*~970 nm with different tilt angles of the incident beam for device of p=505nm. The rotation axis of tilting is along x-axis.*

We further calculate the field profile at resonances. As shown in Fig. S10, for the designed mode at ~970 nm, the electrical field profile inside the device at different locations along the bars aligns with the corresponding eigenmode in Fig. 1. The optical electrical field maintains the same x-polarization as the incident light and propagates along the y-direction. A maximum field enhancement of 16.6-fold is observed at the center of the LNO. It is important to note that the resonance mode retains translational symmetry even with the presence of large notches, without any mode hopping along the propagation direction. This is because most of the mode is confined within the LNO layer with minimal overlap with the notch. This way, the large notches in the bars act as small perturbations to the mode. As shown in Fig. S11, which depicts the simulated transmission for different notch dimensions, we decided not to reduce the notch size to maintain a high resonance contrast, although a Q-factor greater than 1000 is easily achievable with smaller notches. Similar eigenmode and resonance mode analysis was conducted for the other two GMRs, as summarized in Fig. S12. Their distinct mode profiles will affect the EO modulation efficiency, which will be discussed shortly.

In addition to the notch-period-dependent resonance shift, another characteristic of GMR is the resonance splitting induced by oblique incident angles[47,48]. This phenomenon originates from the fact that momentum compensation in GMR relies on the extra momentum gained from the periodic perturbation or grating. For a tilted incidence, the negative and positive orders of diffraction in the medium acquire asymmetrical components along the guided mode propagation direction. As a result, larger or smaller free-space momentum is needed to achieve momentum matching with a tilted incidence, causing a resonance splitting manifested as a blue and a red shift in the free-space wavelengths that would normally excite the same GMR mode. In our device, there are two types of gratings: the super-wavelength free-space grating, composed of a pair of bars repeating along the x-direction as a unit cell, and the subwavelength periodic notches in the y-direction, which diffract light into the LNO and induce GMR along the bars. Since the angle should be referenced relative to the GMR propagation direction, we tilted the device with the rotation axis perpendicular to the bars (x-axis) to observe this effect. The result for the designed resonance at ~970 nm is plotted in Fig. 2 (c), clearly revealing a resonance split upon tilting, which matches well with the simulation in Fig. 2(d). The field profile for the two split resonances, as shown in Fig. S13, indicates they originate from the same mode. The split is increasing with the increase in the tilt angles. At a tilt of 1.5 degrees, a single resonance splits into two, spanning over ~20 nm, suggesting that tilt could be a sensitive tuning mechanism for duplicating a GMR into two resonances at different wavelengths. This resonance splitting with tilt is also observed for other GMR orders at different wavelengths, as shown in Fig. S14.

*Electro-Optic (EO) Tuning*

Following the previous discussion, the device with a notch period of p=505 nm was characterized under a DC bias. To do so, we applied $\pm 10$ V across ITO nanobars and swept the wavelength ranges around the resonances. Higher voltage was not used to avoid any potential (soft) breakdown of the dielectric layers that might interfere with the characteristics. Fig. 3(a) and (b) show the transmission curves of two modes centered at 777.5 nm and 968.5 nm, respectively. The insets provide zoomed-in views of the resonances over a span of 3 nm. The spectrum for the resonance at ~820 nm is shown in Fig. S15. The results clearly display a bidirectional shift in the transmission spectrum under the $\pm 10$ V bias, confirming that the modulation is due to the EO effect. Notably, the amount of lateral shift varies between the different resonances. For $\pm 10$ V, the designed resonance of the fundamental mode at 968.5 nm shows the largest shift of ~0.38 nm; the resonance at 777.5 nm shifts by ~0.2 nm; while the shift for the resonance at ~820 nm is less than 0.1 nm. We verified these experimental results through simulations of the EO effect using a simplified model. The electrical field distribution is first calculated (as shown in Fig. S16). As described in the figure, we divided the LNO layer into several regions and assigned a uniform refractive index change to each region according to the simulated electrical field. For the three resonances mentioned, the simulation (shown in Fig. S17 (a)-(c)) reveals a lateral shift of 0.41 nm (at ~969.5 nm), 0.31 nm (at ~774 nm), 0.06 nm (at ~820 nm), respectively, displaying the same trend as observed in the measurements. The differences in spectral shifts among the modes can be explained with the help of the mode profiles in Fig. S10 and S12. For the simulated resonance at ~774 nm, the mode is also located in the area between the two adjacent bars, similar to the fundamental mode, yet with less mode power inside the LNO (39%) and slightly weaker field enhancement (<14 fold). On the other hand, the simulated resonance at ~820 nm shows a large amount of mode power within the LNO (83%), but the mode spreads across the entire LNO layer, making it difficult to generate a collective EO effect across all the unit cells. Thus, both the spatial distribution and confinement of the mode in the active layer are critical factors for achieving efficient EO tuning. Additionally, as mentioned earlier, we have demonstrated that one mode can split to two when the incident angle deviate from the normal. To this end, we exemplary explored this effect on the fundamental mode for its EO modulation. As expected, the results (shown in Fig. S18) show that both new resonances arising from the split, are capable of providing EO tuning.

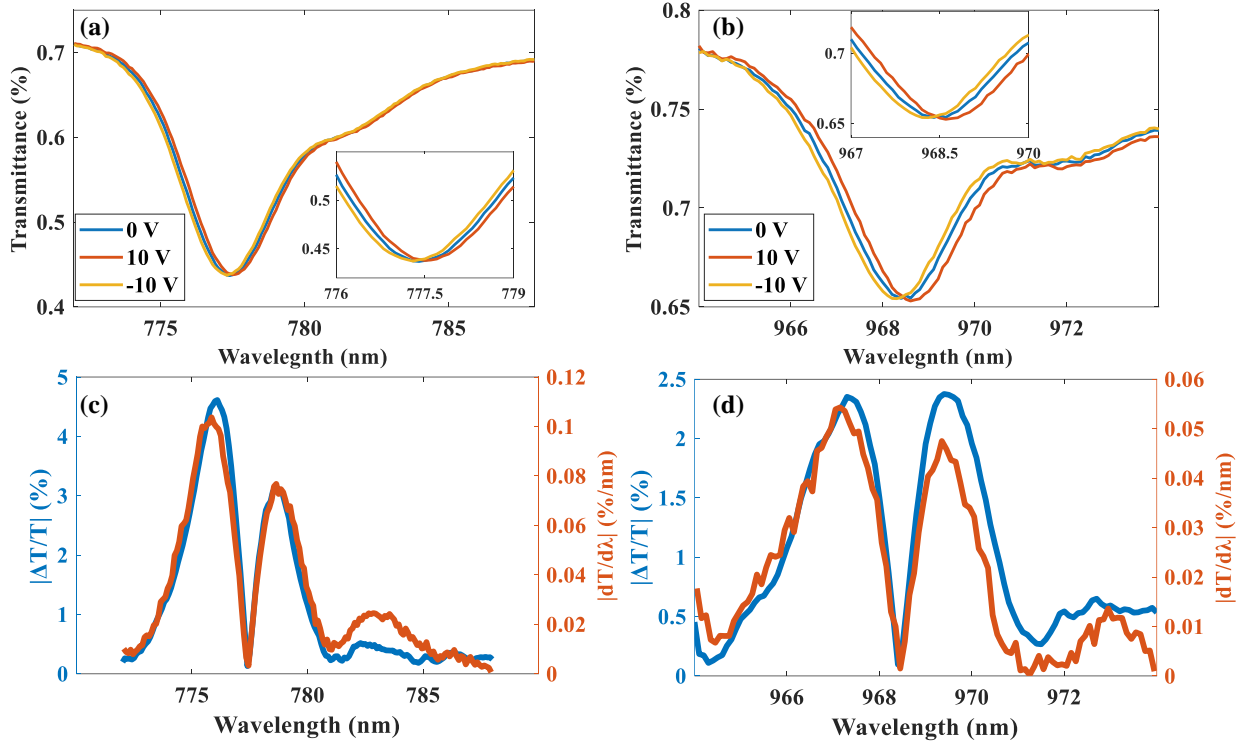

*Fig. 3 Device spectrum shift under DC bias. The spectral shift under $\pm 10$ V of two GMRs at (a) 777.5 nm and (b) the designed fundamental mode at 968.5 nm. The insets show the zoom-in of the spectrum spanning a small range of 3 nm. (c) and (d) The absolute value of modulation ratio (left axis) and the derivative of the transmittance spectrum (right axis) without bias for the two resonances. The modulation ratio is defined as $(T_{10V} - T_{-10V})/T_{0V}$.*

Figure 3 (c), (d) plot the absolute value of modulation ratio ($(T_{10V} - T_{-10V})/T_{0V}$, left axis) and the derivative of transmittance curve (right axis) for the two resonances in (a) and (b), respectively. Clearly, the modulation is minimized when out of resonance. The results also indicate that the larger the derivative, the higher the modulation, with their maximums coinciding nearly at the same wavelength. Under the application of $\pm 10$ V, the highest modulation ratios of 4.6% and 2.35% are achieved at 776.1 nm and 967.3 nm, respectively. Given the similar lateral shifts in both simulation and experiment, larger maximum modulation ratios of 24% and 46.6% are expected from simulation for the resonances centered around 774 nm and 969.5 nm, respectively, as shown in Fig S17 (d)-(e). The simulations and measurements also indicate that the modulation ratios can be affected by both lateral spectrum shift and derivative at resonance slope, where the latter is determined by the Q factor and the contrast of the resonance that can be boosted in experimental device with optimized fabrication process. We therefore believe that a higher modulation ratio is possible to achieve in experiments even given the same lateral spectrum shift as in the current device.

Next, the same device was characterized under AC modulation. Fig. 4 (a) shows the detector response to incident light (right axis) where the device is sinusoidally modulated at a frequency of 100 $Hz$, with a peak-to-peak voltage of $V_{pp} = 20\ V$ (left axis). Without modifying the setup, the laser was set through monochromator at three different wavelengths corresponding to the resonance wavelength (968.5 nm), and two other adjacent wavelengths that yield the highest derivative of the transmission spectrum, on the left (967.4 nm) and right (969.6 nm) sides of the resonance, as depicted in Fig. 3(d). The signal amplitudes are not drawn to scale but are meant to illustrate the trend. The results show that at wavelengths close to either side of the resonance, the detector response follows the modulation frequency: at the left side, the response is in-phase, while at the right side, it is out-of-phase. At resonance, however, the response signal doubles the frequency relative to the modulation signal. We observed similar behavior for other resonances as well (not shown). This behavior is consistent with the DC measurements, confirming that the direction of the transmission curve shift depends on the polarity of the applied field, as a result of EO tuning. During modulations (for various values of $V_{pp}$, including zero), we monitored the DC levels of the response signals and found them to remain constant without a noticeable drift, confirming that thermal effects do not contribute. Figure 4 (b) and (c) present the response to a step function. Following $f_{3dB} = \frac{0.35}{t}$ (where t is rise and fall time), the calculated response frequencies are $f_{rise} = 493\ Hz$ and $f_{rise} = 502\ Hz$. This response frequency is much lower than typical EO devices, and the cause is still under investigation. However, we hypothesize that the non-ideal etching profile of the ITO nanobars' sidewalls may be a contributing factor. The narrow, short bars, especially in regions with notches, likely result in high electrical resistance, increasing the RC time constant and limiting the device's speed.

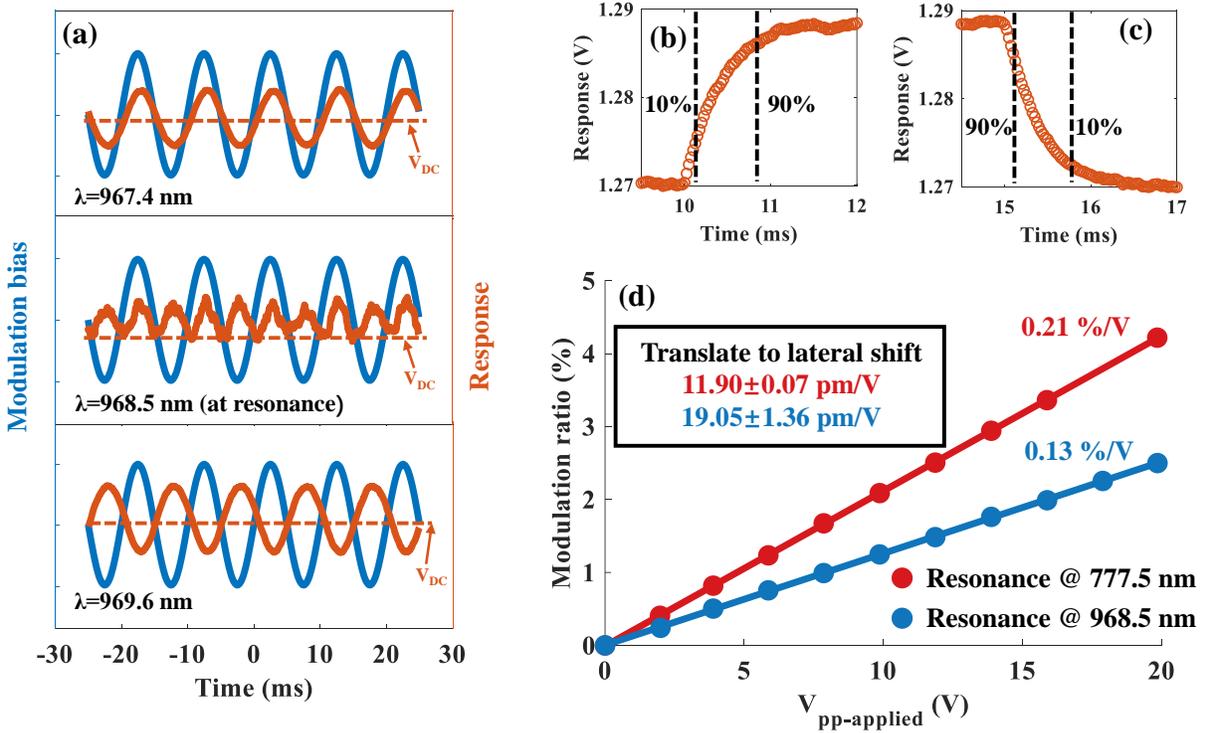

*Fig. 4 Device modulation under AC bias. (a) Transmitted light response (red curves) of the resonance centered at 968.5 nm under 100 Hz sinusoidal modulation (blue curves). The subplots are the responses at different wavelengths around the resonance: at resonance (968.5 nm, middle), maximum derivative points at the right (967.4 nm, top) and left (969.6 nm, bottom) sides of the transmittance curve without bias. The plots are not in scale but show a clear trend that the responses are doubled the frequency (middle), in-phase (top) and out-of-phase (bottom) as to the modulations, respectively. The red dashed lines are the detector signals when with no bias. (b) and (c) Rise and falling edges of the transmitted light response under a 100 Hz squared function modulation. The extracted $f_{3dB}$ are $f_{rise} = 493\ Hz$ and $f_{rise} = 502\ Hz$. (d) The modulation ratios extracted for the resonance at 777.5 nm (red) and 968.5 nm (blue) from different amplitudes of 100 Hz sinusoidal modulations. The probing wavelengths are the maximum derivative points at the right side of each resonance's*

*transmittance curve. The calculated modulation sensitivities are 0.21 %/V and 0.13 %/V, respectively. The inset shows the calculated spectrum lateral shift of each resonance based on the modulation ratio and the transmittance curve derivative. Details can be found in the supporting information.*

Lastly, the device is characterized under varying amplitudes of sinusoidal modulation. The wavelengths were chosen to correspond to the highest derivative point next to the resonance on the right side. The modulation ratio as a function of $V_{pp}$ of modulation signal is outlined in Fig. 4 (d), demonstrating a good linear correlation, consistent with the EO tuning described in equation (1). Modulation sensitivities of 0.21 %/V (4.2% for $\pm 10$ V) and 0.13 %/V (2.6% for $\pm 10$ V) are demonstrated for resonances centered at 777.5 nm and 968.5 nm, respectively. The higher modulation sensitivity for the resonance centered at 777.5 nm is due to its steeper resonance shape that could graphically originate from its higher contrast. Given the linearity of the fits but recognizing that the lateral shifts in the DC measurements were small, we propose a more reliable method for quantifying lateral shifts when a high-resolution tunable laser source is not available. To minimize experimental errors, this method utilizes the largest derivative points, translating the maximum modulation sensitivity into lateral shift in transmittance spectrum, as detailed in Fig. S19 and its followed notes. The inset of Fig. 4 (d) shows the translated lateral shifts of 11.90 pm/V (0.238 nm for $\pm 10$ V) and 19.05 pm/V (0.381 nm for $\pm 10$ V), for resonances at 777.5 nm and 968.5 nm, respectively. Both the modulation sensitivity and the lateral shift agree reasonably well with the results under DC bias, validating this method and confirming the repeatability of device characteristics across different electrical measurements.

**Discussion**

We designed, fabricated and experimentally characterized a transmission type photonic modulator based on the commercially available platform of LNOI. To the best of our knowledge, this is the first experimental demonstration where transparent conductive oxide (TCO) serves both as electrical contacts and in generating resonance. This innovation not only makes the device more compact but also allows greater freedom for optimization due to the significantly lower optical loss of TCO compared to metals[16,18,49]. This feature enables the placement of contacts closer to, or even directly in contact with, the optical mode. Additionally, it simplifies the fabrication process by eliminating the need for extra lithography to align contacts with the metasurface patterns. We demonstrated that, besides creating small perturbation in high-refractive-index waveguides [31,49,50], GMRs are also feasible through periodic notched low-refractive-index TCO with a decent Q-factor and contrast. More importantly, this configuration confines most of the optical mode within LNO when GMR is induced, which enhances EO tuning. For transmission mode operation on LNOI, previous approaches typically placed contacts at two edges of the active area, where local or nonlocal resonance metasurfaces are patterned[21,32–34,37,38,51]. In such schemes, the electric field, which directly influences EO tuning efficiency, depends on the size of the active area, making it difficult to benchmark performance across reported devices. As the device size increases, EO tuning efficiency declines due to the weakened field. Our proposed design addresses this issue by integrating GMR and electrical tuning in each unit cell, allowing all cells to contribute equally to the EO modulation. Furthermore, this configuration still can leverage the largest EO coefficient for efficient modulation.

To boost device performance, several optimization strategies are recommended. First, increasing the distance between adjacent bars allows more of the mode aligned with the electric field to be located between them inside the LNO, though this comes at the expense of a reduction in electric field intensity. Without concern about the optical loss from TCO contacts, this trade-off can be better engineered to enhance EO tuning efficiency. In addition, the super-wavelength unit cell period can be adjusted to create resonance with higher contrast at the $0^{th}$ order diffraction, potentially increasing the modulation amplitude. For example, as shown in Fig. S20, decreasing the period by 200 nm results in ~20% deeper resonance and higher Q factor, with minimal loss of mode confinement within each unit cell. Improving the ITO etching technique can further result in a better etching profile for thicker ITO layers, reducing resistance and speeding up the device response.

In summary, this work demonstrates a highly transmissive free-space light modulator operating in the NIR. The device is based on GMRs induced by transparent conductive metal oxide-ITO meta-grating and thin film LNO without intentional patterning. Through experimental and simulation investigations of EO tuning performance across different modes, we report a designed fundamental resonance at 968.5 nm with a Q factor of 440. This mode, primarily confined within LNO, exhibits the largest resonance shift of 0.38 nm under ±10 V with modulation amplitude of 2.35%. The largest modulation amplitude of 4.6% is observed for another mode with higher contrast at 777.5 nm under the same bias, due to its steeper resonance slope. Additionally, we show that tilting the incident angle provides another degree of freedom to shift and replicate resonances at different wavelengths, both of which can be used for EO modulation. We believe that further optimization will make this device design highly attractive for diverse applications, ranging from active spatial light modulation, optical communications, and sensing to reconfigurable nonlinear applications (due to the field enhancement in LNO), and more.

## Materials and methods

### *Device fabrication*

The device is fabricated from x-cut thin-film LNO (310 nm) on $SiO_2$ (~2 μm) platform, with bulk LNO as substrate (~500 μm), (samples purchased from NANOLN). The fabrication flow of ITO nanobars is schematically shown in Fig. S4. The sample was first cleaned in acetone and piranha. The process started with sputtering (FHR) 100 nm ITO on LNO. Then ZEP-520 resist was spin-coated on ITO followed by ebeam lithography (Elionix) that opened the area of bars. Next, 40 nm Ti was ebeam evaporated (VST) on the patterned resist followed by resist lift-off, so that the bar area was protected by Ti. Consequently, the ITO was patterned by Ar etching of 15 min Ar RIE (Oxford Plasma Lab 100) without ICP to obtain a decent side wall profile[46], and 10 min Ar ion milling (AJA, stage tilt 10°) to etch ITO cleanly between two adjacent bars. We found that the outcome for both Ar etchings under such conditions gave minor over etch to LNO, and the Ti mask was still preserved. Lastly, the residue of Ti mask was removed by 7 min RIE of $CF_4$ and $O_2$ mixed gas. Such recipe can also etch LNO but with very low rate[52], so that the etch profile quality is not a critical consideration here. The overall over-etch is ~50 nm (see zoom-in view in Fig. S5). The active area of ITO meta-grating is 200 μm × 200 μm. After patterning the ITO nanobars, we evaporated 10 nm Cr and 100 nm Au followed by lift-off for electrical pads, patterned using laser writer (Microtech LW405B) on resist LOR5B and AZ1505. The pads were in contact with ITO nanobars at two ends without overlapping with illumination area. The device was finally coated with a layer of PMMA on the top surface, mounted and wire-bonded to a PCB.

*Device characterization*

The optical setup is depicted in Fig. S7. The light source is from a tunable monochromator (Fianium, LLTF VIS-NIR HP8) in connection with a broad band laser (NKT Photonics, SuperK Extreme). The monochromatic light is guided by a fiber which is connected to an adjustable focus fiber collimator (Thorlabs CFC-11X-B). The collimated light is then passed through a linear polarizer and focused down to the device with beam diameter ~90 μm, using a long focal length bi-convex lens (Thorlabs LB1904-B, f=125 mm). An objective (Mitutoyo, M Plan Apo 10X) is used to collect the transmitted light after the sample. Then the light is guided to either a camera (Thorlabs, DCC1545M-GL) through a tube lens, or to a detector (Thorlabs, DET100A/M). The use of long focal length lens is to ensure that the sample is placed around the focal point (ideally plane wave incident) with sufficient tolerance, and the incident cone angle is small if the sample slightly deviates from focal point. By changing the focus of the fiber collimator, we can in principle illuminate the sample at the focal point for different wavelengths. However, we found that if we adjusted the collimator according to some middle point of the wavelength range (e.g. ~875 nm, for 750 nm-1000 nm), the transmittance spectrum curve (Fig. S9) did not show any noticeable difference compared to adjusting for each wavelength range separately (in terms of the resonance center wavelength, Q factor, and contrast). Therefore, it is reasonable to claim that once the collimator is adjusted to the middle point of the sweep range and kept fixed, the effect of the chromatic aberration of the setup on the results is negligible. We also note that the acceptance angle of the objective is smaller than the smallest deviation angle of the $\pm 1$ diffraction orders of the transmitted light from device (NA of objective= $0.28 < sin\theta_{\pm 1} = \frac{\lambda_0}{l} = 0.34$, where $\lambda_0$=750 nm is the lower boundary of the wavelength sweep, $l$ =2200 nm is the unit cell period). Therefore, the objective automatically collects only the $0^{th}$ diffraction order light from the device. For electrical bias, the device is either connected to a DC power source (Keysight, B2901A), or to a function generator (Keysight 3390) for AC modulation.

***Simulation methods***

The eigen mode simulation is conducted using Ansys Lumerical MODE. The periodic boundary condition is assigned in x-direction. The optical transmittance spectrum is simulated by Ansys Lumerical RCWA. COMSOL Multiphysics is used to calculate the electrical field in the device.

## Acknowledgements

The research was partially funded by the Israeli Innovation Authority. Z.C. acknowledges a scholarship from the center for nanoscience and nanotechnology of the Hebrew University. Samples were fabricated at the center for nanoscience and nanotechnology of the Hebrew University.

## Author contributions

Z.C. and U.L. conceived the ideas. N.M. and Z.C. fabricated the device. J.E. and Z.C. suggested and built the optical setup. Z.C. performed the measurements and simulations. U.L. supervised the project.

## Data availability

The data that supports the plots and other findings of this study are available from the corresponding author upon request.

## Conflict of interest

The authors declare no competing interests

# REFERENCES

1. Chen, H.-T., Taylor, A. J. & Yu, N. A review of metasurfaces: physics and applications. *Reports on Progress in Physics* **79**, 076401 (2016).

2. Ding, F., Pors, A. & Bozhevolnyi, S. I. Gradient metasurfaces: a review of fundamentals and applications. *Reports on Progress in Physics* **81**, 026401 (2018).

3. Hsiao, H.-H., Chu, C. H. & Tsai, D. P. Fundamentals and Applications of Metasurfaces. *Small Methods* **1**, 1600064 (2017).

4. Kuznetsov, A. I. *et al.* Roadmap for Optical Metasurfaces. *ACS Photonics* **11**, 816–865 (2024).

5. Gu, T., Kim, H. J., Rivero-Baleine, C. & Hu, J. Reconfigurable metasurfaces towards commercial success. *Nat Photonics* **17**, 48–58 (2023).

6. Jung, C., Lee, E. & Rho, J. The rise of electrically tunable metasurfaces. *Sci Adv* **10**, eado8964 (2024).

7. Zangeneh Kamali, K. *et al.* Electrically programmable solid-state metasurfaces via flash localised heating. *Light Sci Appl* **12**, 40 (2023).

8. Wang, H. *et al.* All-optical ultrafast polarization switching with nonlinear plasmonic metasurfaces. *Sci Adv* **10**, eadk3882 (2024).

9. Klopfer, E., Dagli, S., Barton, D. I. I. I., Lawrence, M. & Dionne, J. A. High-Quality-Factor Silicon-on-Lithium Niobate Metasurfaces for Electro-optically Reconfigurable Wavefront Shaping. *Nano Lett* **22**, 1703–1709 (2022).

10. Sisler, J. *et al.* Electrically tunable space–time metasurfaces at optical frequencies. *Nat Nanotechnol* (2024) doi:10.1038/s41565-024-01728-9.

11. Klopfer, E., Delgado, H. C., Dagli, S., Lawrence, M. & Dionne, J. A. A thermally controlled high-Q metasurface lens. *Appl Phys Lett* **122**, 221701 (2023).

12. Damgaard-Carstensen, C., Thomaschewski, M., Ding, F. & Bozhevolnyi, S. I. Electrical Tuning of Fresnel Lens in Reflection. *ACS Photonics* **8**, 1576–1581 (2021).

13. Li, J., Chen, Y., Hu, Y., Duan, H. & Liu, N. Magnesium-Based Metasurfaces for Dual-Function Switching between Dynamic Holography and Dynamic Color Display. *ACS Nano* **14**, 7892–7898 (2020).

14. Han, Z., Frydendahl, C., Mazurski, N. & Levy, U. *MEMS Cantilever-Controlled Plasmonic Colors for Sustainable Optical Displays*. *Sci. Adv* vol. 8 https://www.science.org (2022).

15. Park, J. *et al.* All-solid-state spatial light modulator with independent phase and amplitude control for three-dimensional LiDAR applications. *Nat Nanotechnol* **16**, 69–76 (2021).

16. Benea-Chelmus, I.-C. *et al.* Gigahertz free-space electro-optic modulators based on Mie resonances. *Nat Commun* **13**, 3170 (2022).

17. Zhang, J. *et al.* High-speed metasurface modulator using perfectly absorptive bimodal plasmonic resonance. *APL Photonics* **8**, 121304 (2023).

18. Zheng, T., Gu, Y., Kwon, H., Roberts, G. & Faraon, A. Dynamic light manipulation via silicon-organic slot metasurfaces. *Nat Commun* **15**, 1557 (2024).

19. Weigand, H. C. *et al.* Nanoimprinting Solution-Derived Barium Titanate for Electro-Optic Metasurfaces. *Nano Lett* **24**, 5536–5542 (2024).

20. Jeon, N., Noh, J., Jung, C. & Rho, J. Electrically tunable metasurfaces: from direct to indirect mechanisms. *New J Phys* **24**, 075001 (2022).

21. Trajtenberg-Mills, S. *et al.* LNoS: Lithium Niobate on Silicon Spatial Light Modulator. *ArXiv* (2024).

22. Koeber, S. *et al.* Femtojoule electro-optic modulation using a silicon–organic hybrid device. *Light Sci Appl* **4**, e255–e255 (2015).

23. Zhang, J. *et al.* High-speed metasurface modulator using perfectly absorptive bimodal plasmonic resonance. *APL Photonics* **8**, 121304 (2023).

24. Eppenberger, M. *et al.* Resonant plasmonic micro-racetrack modulators with high bandwidth and high temperature tolerance. *Nat Photonics* **17**, (2023).

25. Shur, V. Ya., Akhmatkhanov, A. R. & Baturin, I. S. Micro- and nano-domain engineering in lithium niobate. *Appl Phys Rev* **2**, 040604 (2015).

26. Boes, A. *et al.* Lithium niobate photonics: Unlocking the electromagnetic spectrum. *Science (1979)* **379**, eabj4396 (2024).

27. Fedotova, A. *et al.* Lithium Niobate Meta-Optics. *ACS Photonics* **9**, 3745–3763 (2022).

28. You, B. *et al.* Lithium niobate on insulator – fundamental opto-electronic properties and photonic device prospects. *Nanophotonics* **13**, 3037–3057 (2024).

29. Chen, G., Gao, Y., Lin, H.-L. & Danner, A. J. Compact and Efficient Thin-Film Lithium Niobate Modulators. *Adv Photonics Res* **4**, 2300229 (2023).

30. Zhu, D. *et al.* Integrated photonics on thin-film lithium niobate. *Adv Opt Photonics* **13**, 242–352 (2021).

31. Klopfer, E., Dagli, S., Barton, D. I. I. I., Lawrence, M. & Dionne, J. A. High-Quality-Factor Silicon-on-Lithium Niobate Metasurfaces for Electro-optically Reconfigurable Wavefront Shaping. *Nano Lett* **22**, 1703–1709 (2022).

32. Gao, B., Ren, M., Wu, W., Cai, W. & Xu, J. Electro-optic lithium niobate metasurfaces. *Sci China Phys Mech Astron* **64**, 240362 (2021).

33. Weigand, H. *et al.* Enhanced Electro-Optic Modulation in Resonant Metasurfaces of Lithium Niobate. *ACS Photonics* **8**, 3004–3009 (2021).

34. Ju, Y. *et al.* Hybrid resonance metasurface for a lithium niobate electro-optical modulator. *Opt Lett* **47**, 5905–5908 (2022).

35. Damgaard-Carstensen, C. & Bozhevolnyi, S. I. Nonlocal electro-optic metasurfaces for free-space light modulation. *Nanophotonics* **12**, 2953–2962 (2023).

36. Liu, J. *et al.* Lithium niobate thin film electro-optic modulator. *Nanophotonics* **13**, 1503–1508 (2024).

37. Ju, Y. *et al.* The electro-optic spatial light modulator of lithium niobate metasurface based on plasmonic quasi-bound states in the continuum. *Nanoscale* **15**, 13965–13970 (2023).

38. Ju, Y., Zhang, W., Zhao, Y., Deng, X. & Zuo, H. Polarization independent lithium niobate electro-optic modulator based on guided mode resonance. *Opt Mater (Amst)* **148**, 114928 (2024).

39. Weiss, A. *et al.* Tunable Metasurface Using Thin-Film Lithium Niobate in the Telecom Regime. *ACS Photonics* **9**, 605–612 (2022).

40. Chen, G., Cheung, E. J. H., Cao, Y., Pan, J. & Danner, A. J. Analysis of perovskite oxide etching using argon inductively coupled plasmas for photonics applications. *Nanoscale Res Lett* **16**, 32 (2021).

41. Mookherjea, S., Mere, V. & Valdez, F. Thin-film lithium niobate electro-optic modulators: To etch or not to etch. *Appl Phys Lett* **122**, 120501 (2023).

42. Leng, R., Chen, X., Liu, P., Zhu, Z. & Zhang, J. High Q lithium niobate metasurfaces with transparent electrodes for efficient amplitude and phase modulation. *Appl Opt* **63**, 3156–3161 (2024).

43. Liu, G., Zong, S., Liu, X., Chen, J. & Liu, Z. High-performance etchless lithium niobate layer electro-optic modulator enabled by quasi-BICs. *Opt Lett* **49**, 113–116 (2024).

44. Zhang, M., Wang, C., Kharel, P., Zhu, D. & Lončar, M. Integrated lithium niobate electro-optic modulators: when performance meets scalability. *Optica* **8**, 652–667 (2021).

45. Lawrence, M. *et al.* High quality factor phase gradient metasurfaces. *Nat Nanotechnol* **15**, 956–961 (2020).

46. Han, K. *et al.* Indium-Gallium-Zinc-Oxide (IGZO) Nanowire Transistors. *IEEE Trans Electron Devices* **68**, 6610–6616 (2021).

47. Wang, S. S. & Magnusson, R. Theory and applications of guided-mode resonance filters. *Appl Opt* **32**, 2606–2613 (1993).

48. Valencia Molina, L. *et al.* Enhanced Infrared Vision by Nonlinear Up-Conversion in Nonlocal Metasurfaces. *Advanced Materials* **36**, 2402777 (2024).

49. Dagli, S. *et al.* Free-space electro-optic modulators using high quality factor silicon on lithium niobate metasurfaces. in *CLEO 2024* FM4O.5 (Optica Publishing Group, Charlotte, North Carolina, 2024). doi:10.1364/CLEO_FS.2024.FM4O.5.

50. Barton III, D., Lawrence, M. & Dionne, J. Wavefront shaping and modulation with resonant electro-optic phase gradient metasurfaces. *Appl Phys Lett* **118**, 071104 (2021).

51. Chen, X. *et al.* High quality factor resonant metasurface with etchless lithium niobate. *Opt Laser Technol* **161**, 109163 (2023).

52. Winnall, S. & Winderbaum, S. *Lithium Niobate Reactive Ion Etching*. (DEFENCE SCIENCE AND TECHNOLOGY ORGANIZATION SALISBURY (AUSTRALIA), 2000).

# Supporting information

# Tunable transmissive metasurface based on thin-film lithium niobate

*Zetian Chen, Noa Mazurski, Jacob Engelberg, and Uriel Levy*

*Corresponding authors: Zetian Chen and Uriel Levy*

Institute of Applied Physics, The Faculty of Science, The Center for Nanoscience and Nanotechnology, The Hebrew University of Jerusalem

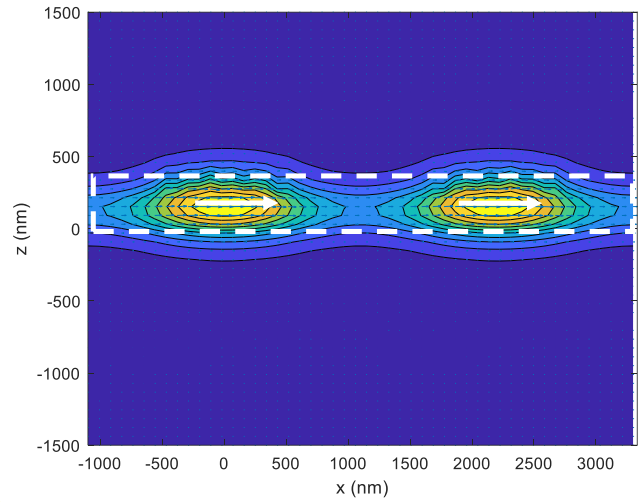

*Fig. S1 vector plot of the eigen mode shown in Fig. 1 (b), for two unit cells. It shows TE (x) polarized mode. The area of white dashed line represents the LNO.*

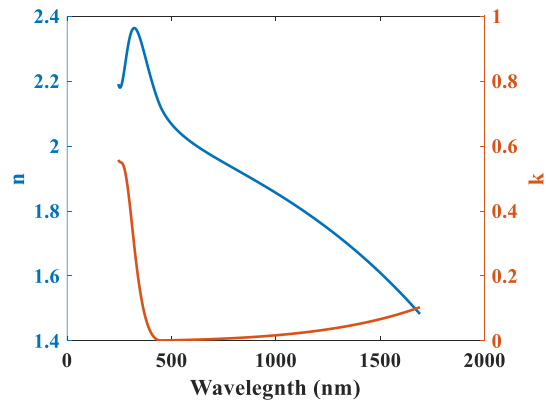

*Fig. S2 n and k of the ITO used in this work, as measured by ellipsometer. The ITO shows low k below 1 um.*

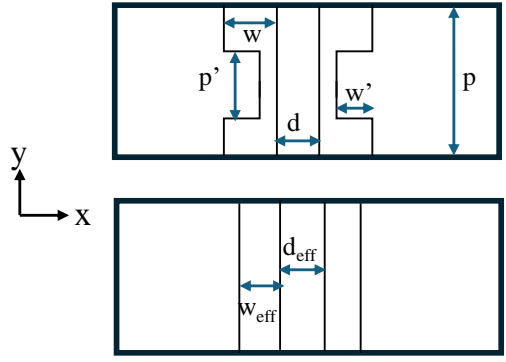

*Fig. S3 Top view of a single period along y for one unit cell. Top figure is the original device with periodic notch along bars. Bottom figure is a translated symmetrical structure without notch, which is used for engine mode and electrical field calculations.*

To simplify the analysis of eigen mode and electrical field calculations, we translate notched structure into a translational symmetrical structure along y without notch. The relationship between the parameters is as following:

$$d_{eff} = d$$

$$w_{eff} = w \cdot \left(1 - \frac{p'}{p}\right) + w \cdot \left(1 - \frac{w'}{w}\right) \cdot \frac{p'}{p} = w \cdot \left(1 - \frac{p'}{p} \cdot \frac{w'}{w}\right)$$

where w' and p' are the notch dimensions, w and p are the bar width without notch and the notch period, d is the bar distance, and $d_{eff}$ and $w_{eff}$ are the bars distance and bar width after transformation. Given the dimension in the main text, the $w_{eff}$=258.4 nm. The idea of transformation is simply keeping the bar distance unchanged, while average the bar width along the y direction. We believe that the transformed structure captures the significance of electrical field in notched device, as the largest field depends on the closest bar distance. As also mentioned in the main text later, the feasibility of such a translation relies on the fact that even though the notch dimension is large, the mode is primarily inside seemingly unpatterned LNO.

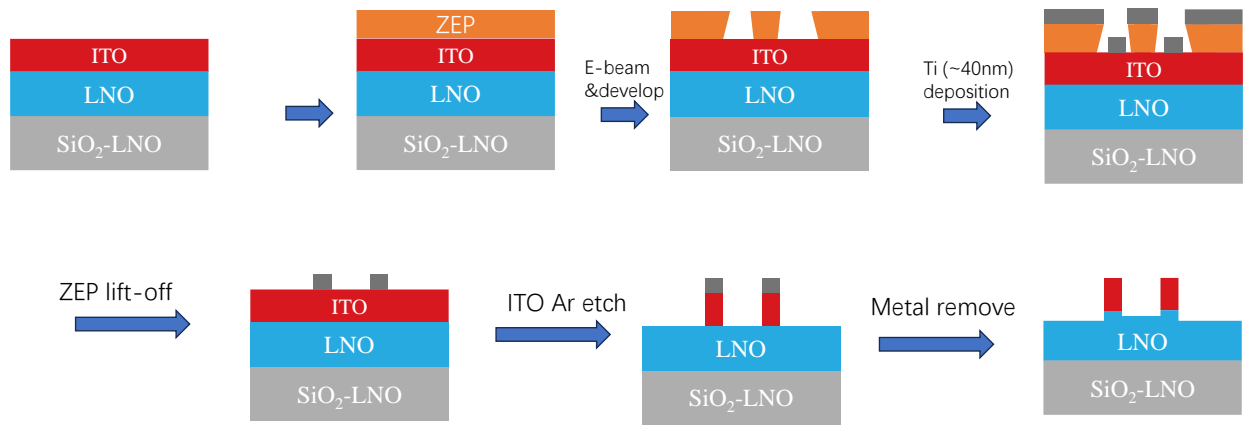

Fig. S4 The fabrication flow of patterning the ITO nanobars.

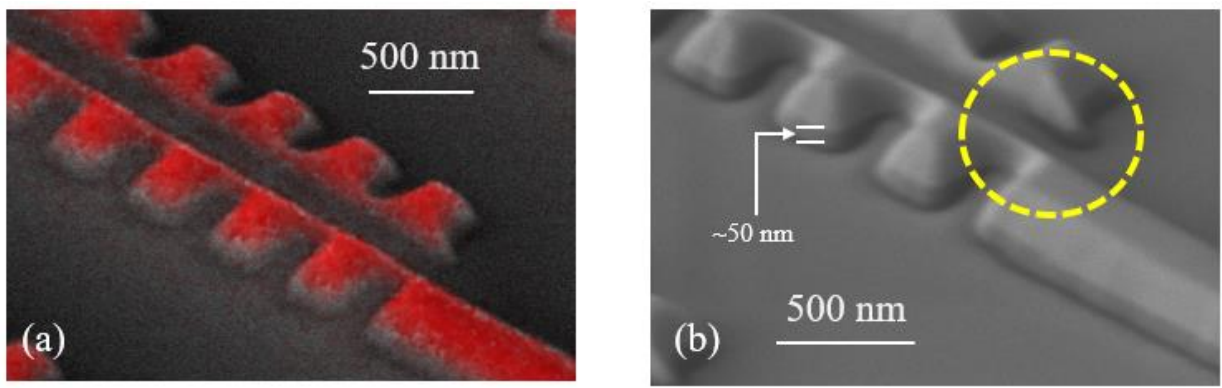

Fig. S5 (a) EDS Indium element mapping shows clean etching of ITO between the two bars. (b) Zoom-in SEM image with tilted 55°. At outside of the notched bars, a ~50 nm over etch into LNO can be seen. The yellow dashed circle shows that the area between the bars is slightly less over etched than the surroundings. We believe it is because of a micro-loading effect during the RIE process, resulting in the area with narrow gaps to be etched slower than the open area.

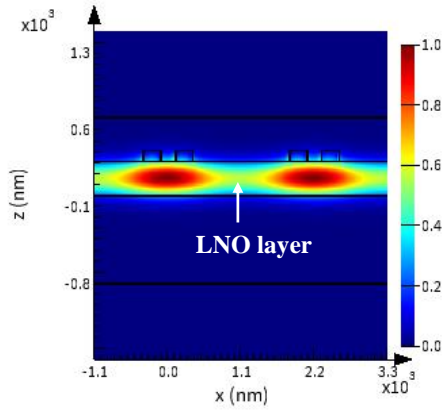

*Fig. S6. Eigen mode electrical field intensity profile of two neighboring unit cells for a structure without any over etch to the LNO. Compared to Fig. 1 (b), the mode here is less confined within each unit cell.*

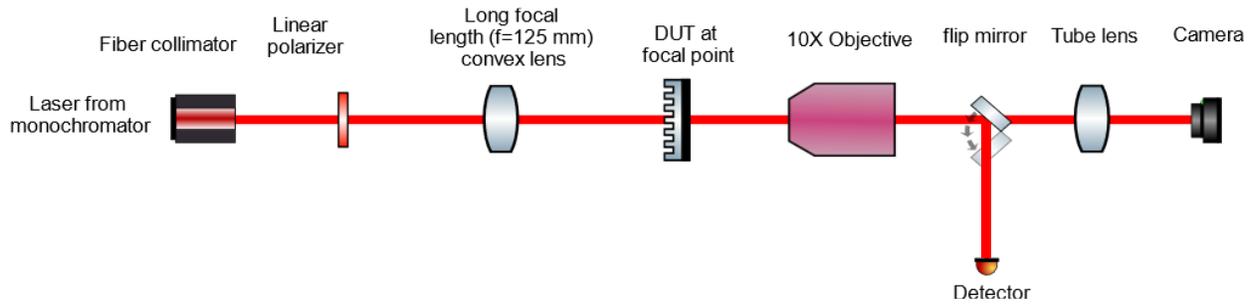

*Fig. S7 Optical measurement setup.*

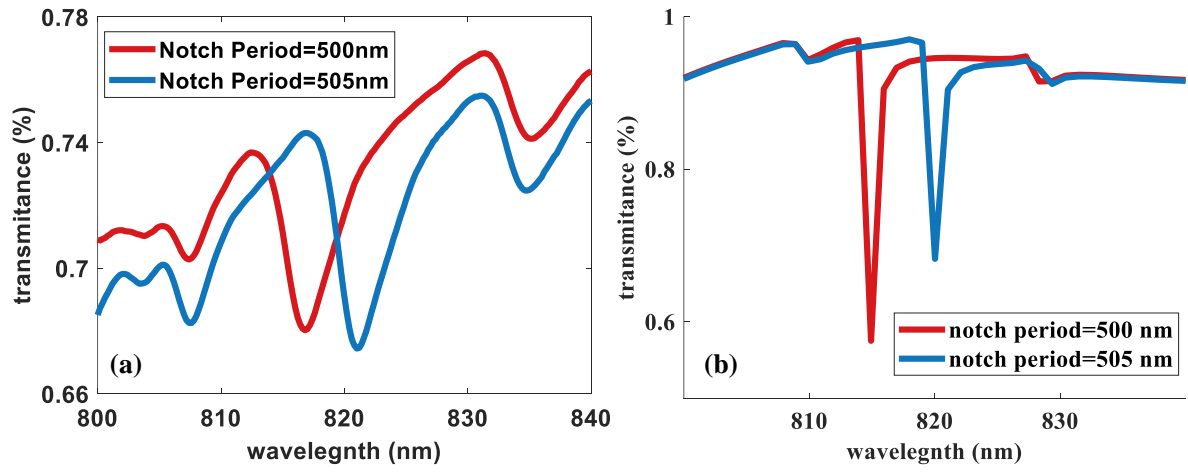

*Fig. S8 (a) Measured and (b) simulated spectrum for the resonance at ~820 nm.*

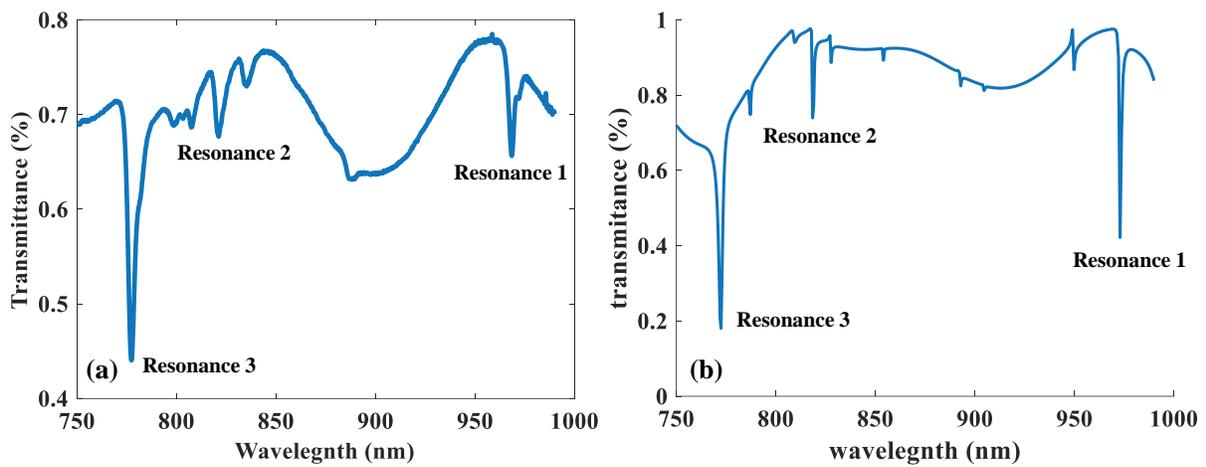

*Fig. S9 (a) Measured and (b) simulated transmission spectrum from 750 nm-990 nm of device with notch period=505 nm. There are three most prominent resonances that are in good match between measurement and simulation: ~970 nm, ~820 nm, ~770 nm. The device is measured in a single wavelength sweep with fixed fiber collimator focal length.*

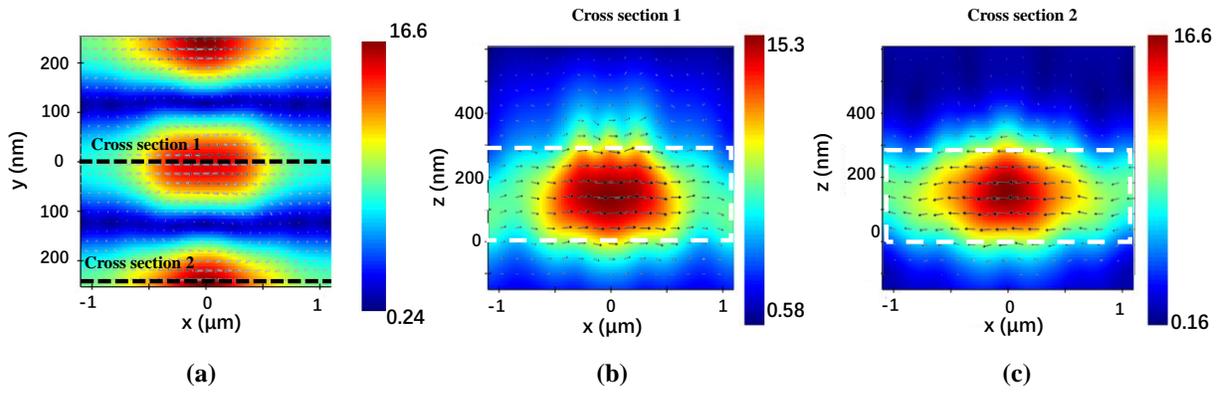

*Fig. S10 Calculated optical E field profile at resonance for the mode at ~970 nm. (a) Field in the xy-plane in the middle of LNO layer. (b) Field in the xz-plane in the middle of the notch. (c) Field in the xz-plane in the middle of the unnotched region.*

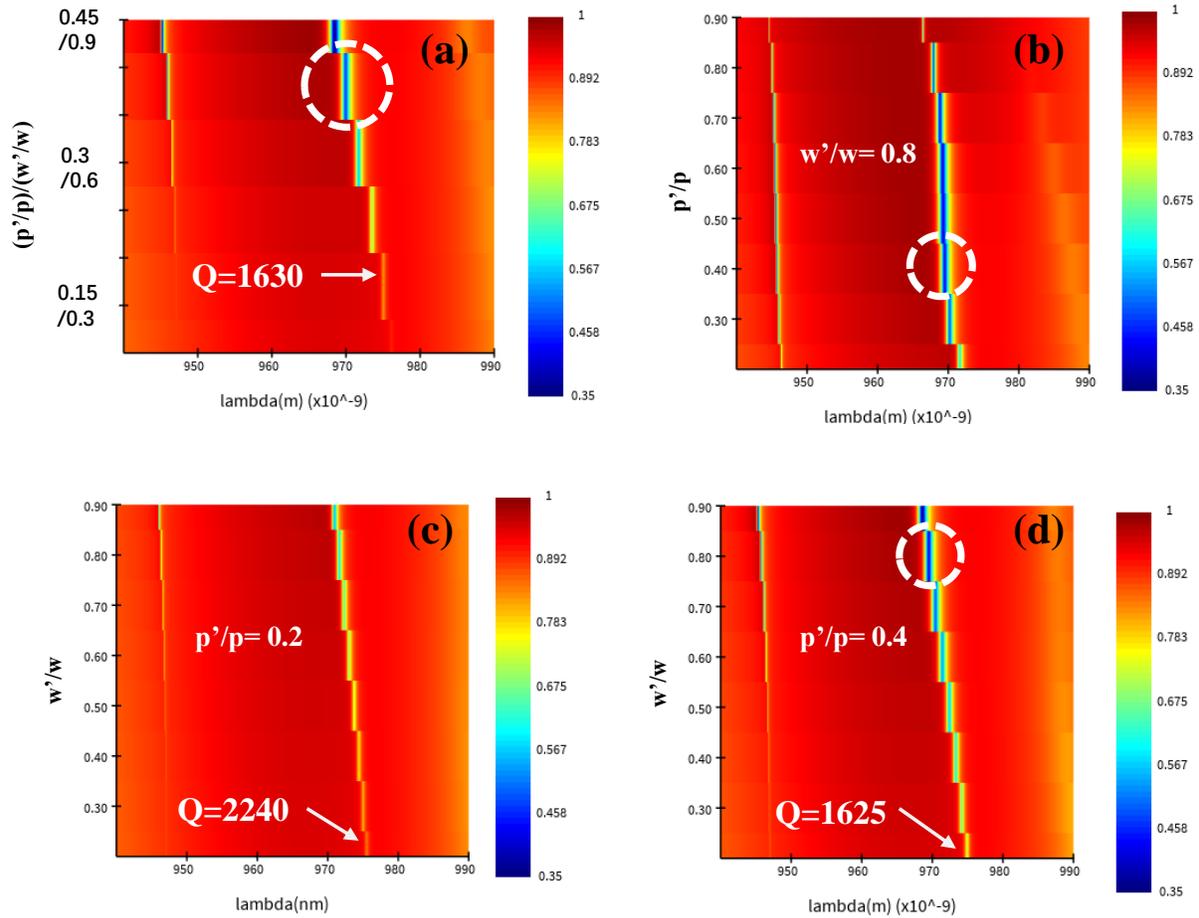

Fig. S11 Simulated transmission spectrum for different notch dimensions. Definition of w, w', p, p' can be referred to in Fig. 1 (c). (a) Keep the notch length and width ratio fixed to be 1:2 and sweep the notch size. (b) Notch length sweep with w'/w=0.8. (c) Notch width sweep with p'/p=0.2. (d) Notch width sweep with p'/p=0.4. The color bars are in the same scale for better comparison. The white dashed circle shows the dimension used in the reported device. Higher Q-factor can be obtained for smaller notch but at the cost of losing resonance contrast.

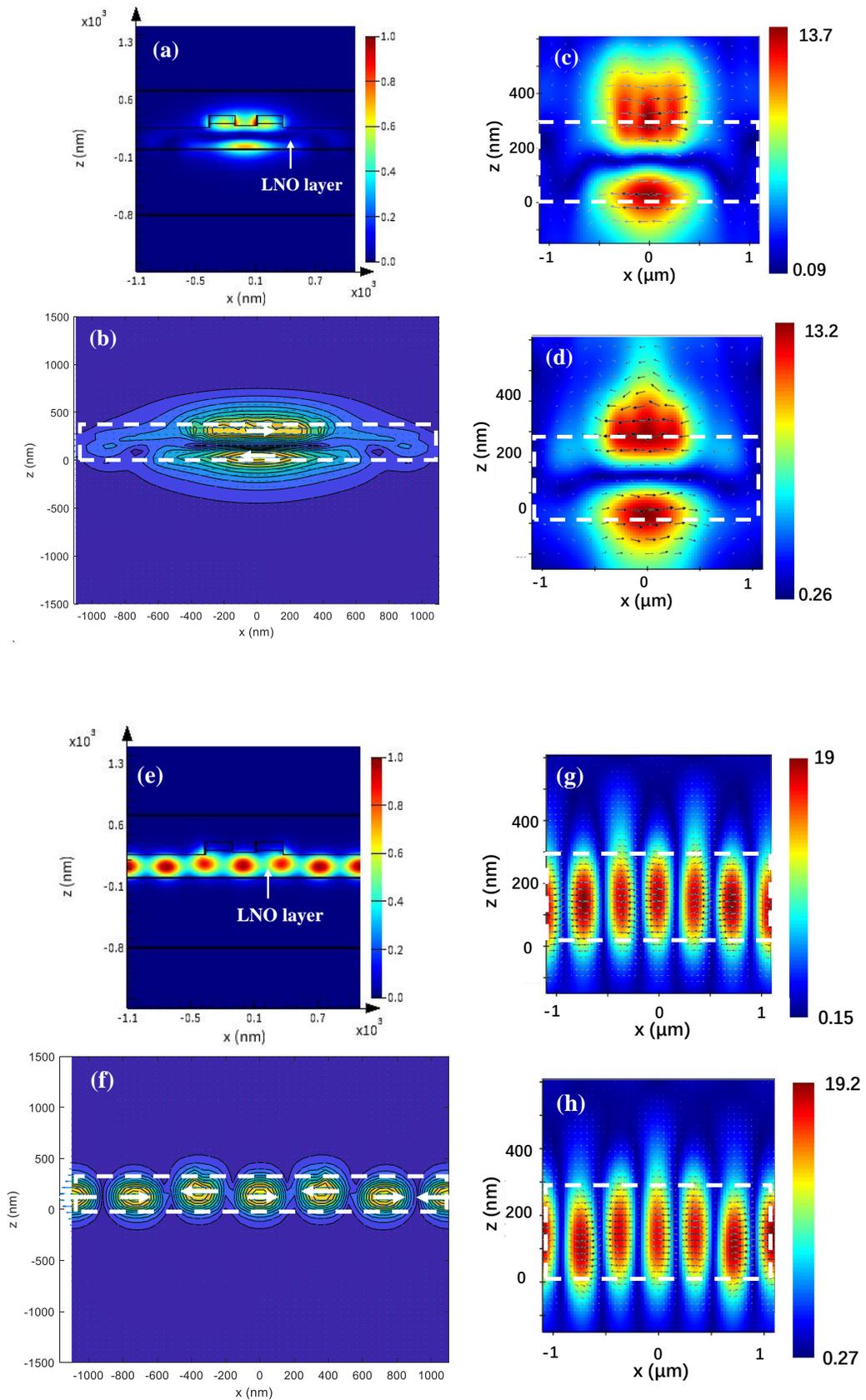

*Fig. S12 Eigen mode analysis and optical E field calculated at resonances, for the GMRs at (a)-(d) ~775 nm, and (e)-(f) ~820 nm. (a) and (e) Eigen mode electrical field intensity profiles. (b) and (f) The corresponding field vector plot. The optical E fields are monitored in the xz-plane at the middle of the notch ((c) and (g)), and at the middle of the un-notched region ((d) and (h)).*

Like the resonance at ~970 nm, the results in Fig. S12 show good match between eigne mode electrical field profile and electrical field profile at resonances. The calculated mode effective refractive index in (a) and (e) are 1.55 and 1.624, respectively. According to equation. (2) in the main text, they are in line with the wavelengths that the two resonances are supposed to appear (~775 nm and ~820 nm, given notch period p=505 nm).

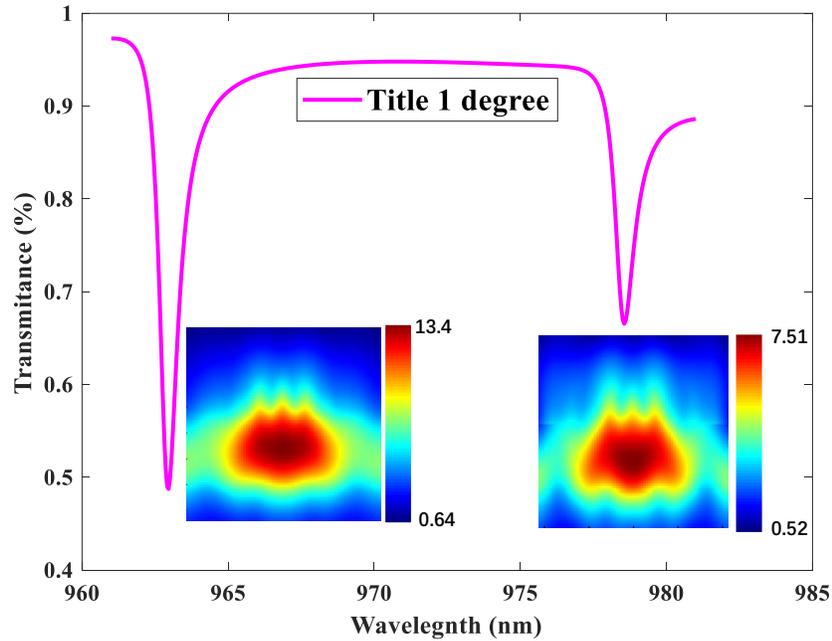

*Fig. S13 The optical E field for the two new resonances upon 1° tilt. They show a similar profile as for the fundamental mode.*

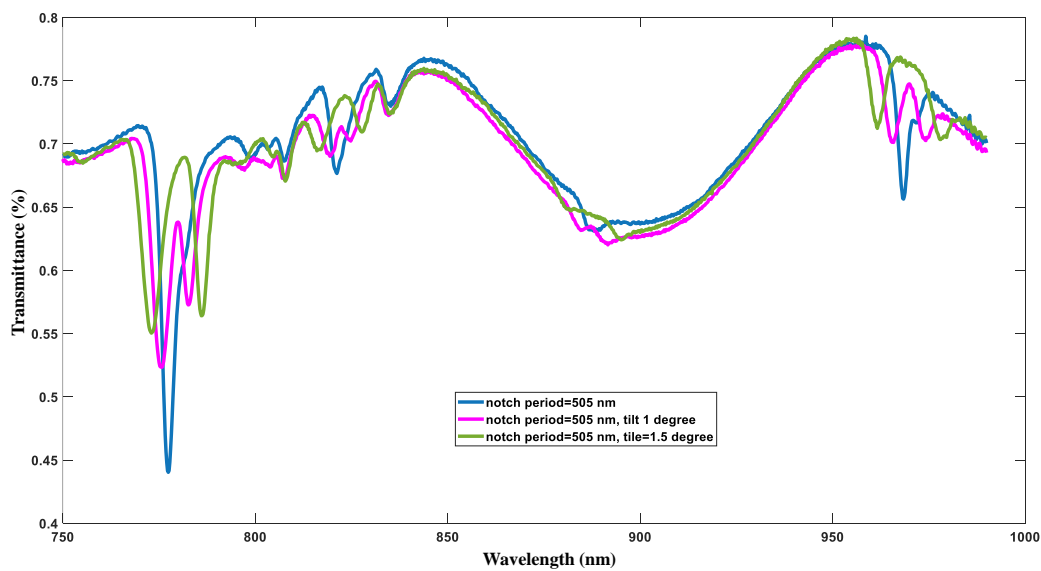

*Fig. S14 Measured transmittance spectrum from 750 nm-990 nm upon tilt. The split of resonance also happens for the resonances at ~775 nm and ~820 nm.*

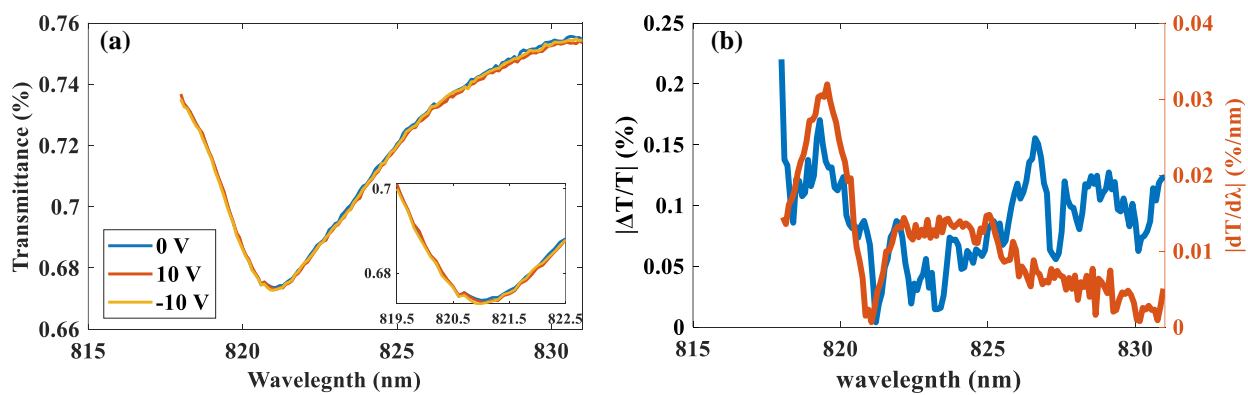

*Fig. S15 Device spectrum modulation under DC bias for the resonance at ~820 nm. (a) Spectrum taken under ±10 V bias. The insets show the zoom-in of the spectrum spanning a small range of 3 nm. (b) The absolute value of modulation ratio (left axis) and the derivative of the transmittance spectrum (right axis) without bias*

*for the two resonances. The modulation ratio is defined as $(T_{10V}-T_{-10V})/T_{0V}$. The resonance shift is very small (<0.1 nm) for this resonance.*

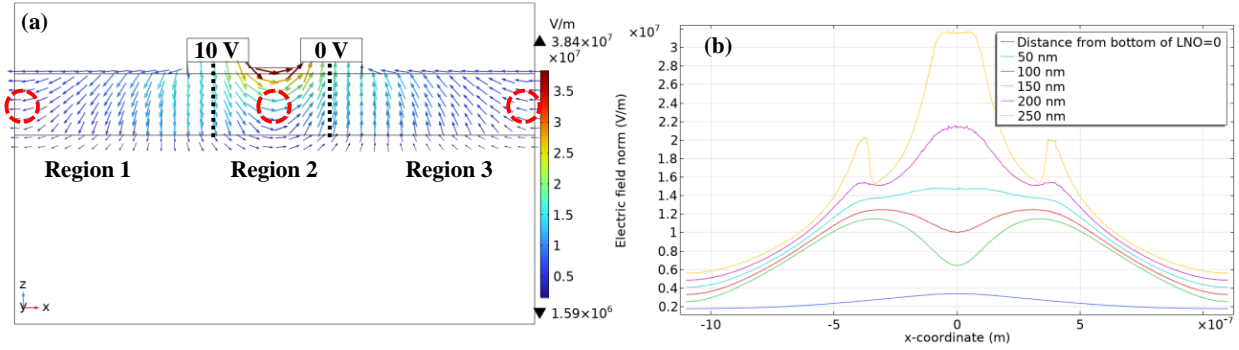

*Fig. S16 Electrical field simulation within one unit cell. Periodic boundary conditions are assigned to the left and right sides of the cell. (a) Electrical field vector plot (only the field inside LNO layer is shown). (b) The electrical field amplitude at different distance from the bottom of the LNO.*

To simulate the EO effect through a simplified model, we first divide LNO layer into region 1, 2 and 3. The boundaries of region 2 are the middle of each bar (after the transformed structure in Fig. S3). We use the electrical field at 150 nm (around the middle) from the bottom of LNO (shallow blue line in (b)) to represent the field intensity of each region. We further assign a single field value, which is the place where the field is parallel with x-axis, to calculate the refractive index change based on equation. (1) in the main text. The corresponding places in the device are marked as red circles in (a). The value used in the simulation are 1.5e7 V/m (middle) and -0.4e7 V/m (sides).

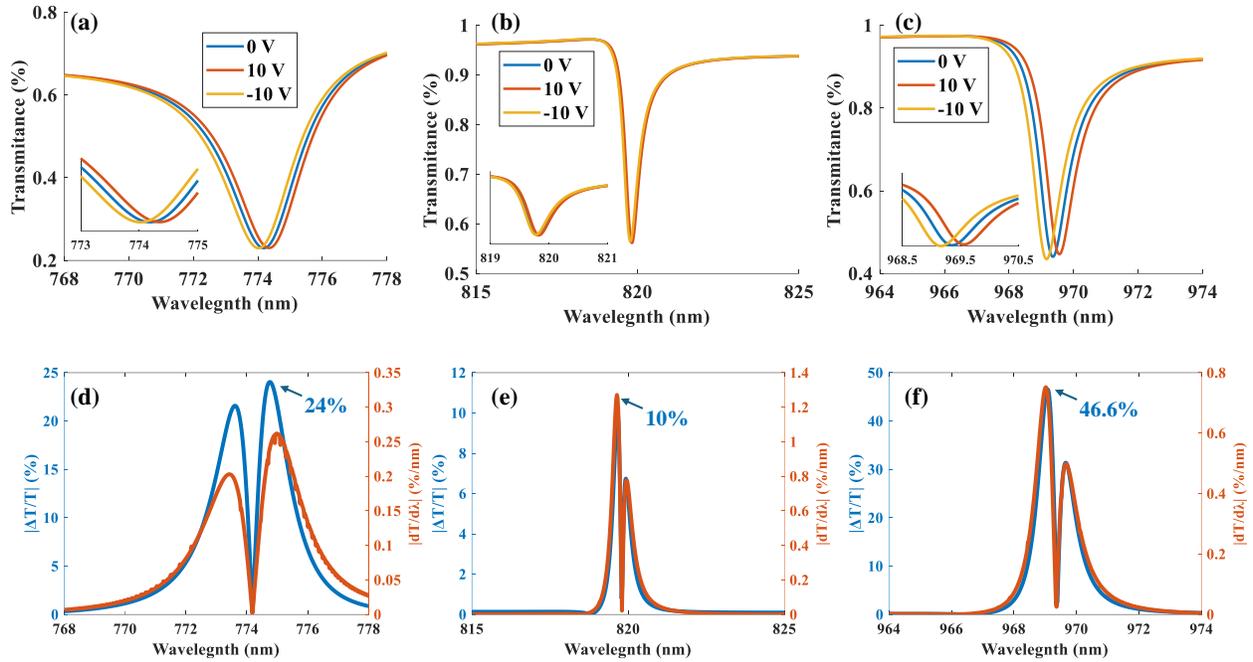

*Fig. S17 Simulated spectral shift under ±10 V DC bias for the three resonances in the simulation centered (a) 774 nm, (b) 820 nm, and (c) 969.5 nm. The insets show the zoom-in of the spectrum spanning a small range of 2 nm. The extracted spectrum lateral shifts are (a) 0.31 nm, (b) 0.06 nm, and (c) 0.41 nm. (d)-(f) The absolute value of modulation ratio (right axis) and the derivative of the transmittance spectrum without bias for the three resonances. The modulation ratio is defined as $(T_{10V}-T_{-10V})/T_{0V}$. Given the similar amount of lateral shift, the modulation ratio is higher in the simulation than measurements because of the higher simulated Q-factor giving steeper slop around the resonances. The resonance at ~820 nm is slightly deeper than in Fig. S8 (b), since we used a finer wavelength sweep here to better capture the small resonance shift.*

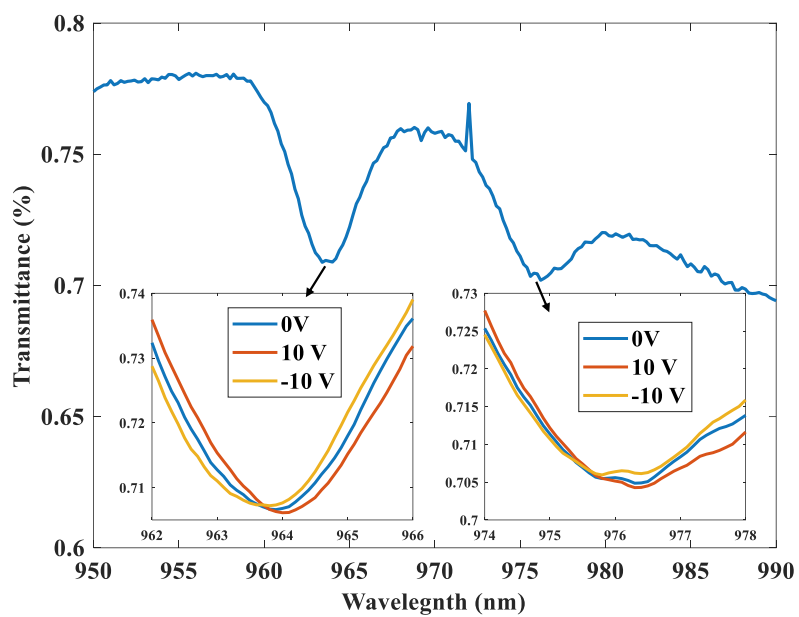

*Fig. S18 The measured spectrum under ±10 V DC bias for the fundamental mode resonance upon tilt. The tilt angle here is ~1.2°.*

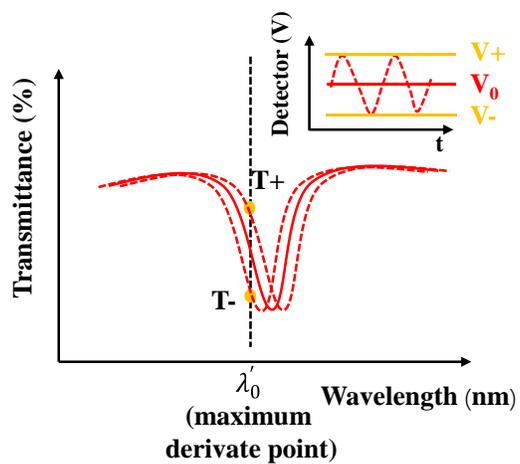

*Fig. S19 Schematic of the spectral shift upon the application of a DC bias. The inset shows the corresponding AC modulation response under a sinusoid function at the probing wavelength (dashed line).*

We assume a small lateral shift to the spectrum around the resonance upon DC bias. Then the derivative of the transmittance curve at certain wavelength on the slop can be approximated as:

$$derivate := |\tan(\theta)|\,[\%/nm] = \frac{T_+ - T_-}{\Delta\lambda}$$

where $T_+$ and $T_-$ are the transmittance (in %) under positive and negative bias at the observe wavelength, $\Delta\lambda$ is the spectrum shift. The spectrum is measured using detector normalized to the background (air) response:

$$T(\lambda) = V(\lambda)/V_{bg}(\lambda)$$

where $V_{bg}(\lambda)$ is the background detector readout, and $V(\lambda)$ is the signal with device. Therefore, the lateral shift can be represented as:

$$\Delta\lambda[nm] = \frac{T_+ - T_-}{|\tan(\theta)|\,[\%/nm]} = \frac{V_+/V_{bg} - V_-/V_{bg}}{|\tan(\theta)|} = \frac{V_+ - V_-}{|\tan(\theta)| \times V_{bg}}$$

where $V_+$ and $V_-$ is the detector read out under positive and negative bias. They can be read also from the response upon the sinusoid modulation, shown in the inset of the Fig. S19. The shift can be further normalized to the amplitude ($V_{pp}$) of the modulation:

$$\frac{\Delta\lambda}{\Delta V_{bias}}\left[\frac{nm}{V}\right] = \frac{(V_+ - V_-)/V_{pp}}{|\tan(\theta)| \times V_{bg}} = \frac{(V_+ - V_-)/V_{pp}}{V_0} \cdot \frac{V_0(\lambda)}{V_{bg}(\lambda)} \cdot \frac{1}{|\tan(\theta)|}$$

$$= (modulation\ senstivity) \cdot T_0(\lambda) \cdot \frac{1}{|\tan(\theta)|}$$

where $V_0$ is the detector readout without bias, and $modulation\ senstivity\ [\%/V] := \frac{(V_+ - V_-)/V_{pp}}{V_0}$ is measured in Fig. 4 (d). Accordingly, the lateral shift can be extracted from different amplitudes of AC modulation as well, as shown in the inset of the Fig. 4 (d). The tolerance comes from the two maximum derivative points at two sides of a resonance that is used to extract the results.

Upon bias, we found, both from measurement (Fig. 3) and simulation (Fig. S17), that the spectrum still shows a seemingly parallel shift within a certain range of wavelength, including at least the maximum derivative points. Therefore, it is reasonable to claim that evaluating from the maximum derivative points can truly reflect the resonance shift. Using maximum derivative also helps to reduce the calculation error since the derivative is in the denominator, while in principle we can obtain the same shift by examining other derivative points within the parallel shift wavelength range. We believe that extracting resonance shift from AC modulation as described is in general more accurate and robust than reading directly from DC induced shift. This is because averaging and normalization are considered during calculations. For example, for each amplitude of AC modulation, the modulation ratio is averaged over multiple periods. Additionally, if assume detector response and modulation amplitude are linear within a small range of incident power, the method proposed here does not require an ultra-stable source during several wavelength sweeps, since the modulation ratio is normalized to a new $V_0$ every time the AC modulation amplitude changed, and $T_0(\lambda)$ is a fixed value from a single sweep. In contrast, small fluctuations on the source power will slightly move the transmittance curve up or down during wavelength sweeps, introducing artifacts when read shift directly from multiple DC-biased wavelength sweeps. This is increasingly important when the lateral shift is small to begin with.

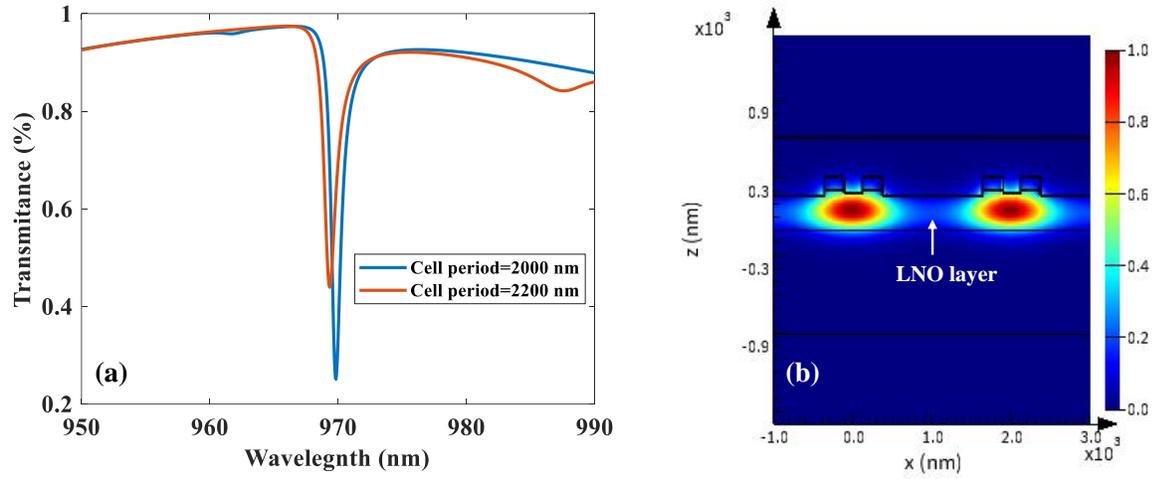

*Fig. S20 Simulated spectrum (a) and eigen mode of two unit cells (b) for the fundamental mode resonance with cell period of 2000 nm. It shows a higher contrast and higher Q-factor (1080) than the cell period of 2200 nm. The mode is still reasonably confined within each unit cell, compared to Fig. 1 (b).*